\documentclass[twoside]{article}

\usepackage{amsmath}
\usepackage{amssymb}
\usepackage{pstricks}
\usepackage{epsfig}
\usepackage{fleqn,espcrc2}

\newcommand{\mb}{\mathbf}
\newcommand{\Astat}{A^{\text{stat}}_0}
\newcommand{\Argi}{(A^{\text{RGI}})_0}
\newcommand{\zastat}{Z_{\rm A}^{\text{stat}}}
\newcommand{\Aren}{(A^{\text{stat}}_{\rm R})_0}
\newcommand{\heavy}{\psi_{\rm h}}
\newcommand{\heavyb}{\bar{\psi}_{\rm h}}

\title{Renormalization of the static-light axial current}

\author{Martin Kurth and Rainer Sommer\address{DESY Zeuthen,
    Platanenallee 6, 15738 Zeuthen, Germany}}

\begin{document}

\begin{abstract}
We discuss the determination of the heavy-light
axial current renormalization in the static approximation,
using a new method based on the Schr\"odinger Functional (SF).
Previous perturbative results
for the renormalization constant are confirmed.
\end{abstract}

\maketitle
\section{Introduction and strategy}
  The  static limit is of 
  interest in the computation of $f_{\rm B}$ and the 
  control of the systematic errors involved. In order
  to obtain a truly
  non-perturbative answer for $f_{\rm B}$ in that limit, one first
  needs to solve  the
  renormalization 
  problem for the static
  axial current. This is of practical importance
  as underlined in S.~Hashimoto's and Y.~Kuramashi's
  reviews~\cite{HASHIMOTO1,KURAMASHI1}.
 
  The ALPHA-collaboration developed a strategy for non-perturbative
  renormalization and applied it to the coupling and the quark masses
  of QCD~\cite{CAPITANI2}. The same strategy applies to the static axial
  current. The final goal in this case is to compute the
  renormalization constant relating the bare current in the lattice
  regularization to the renormalization group invariant (RGI) current 
\begin{eqnarray*}
  \Argi  \propto \lim_{\mu\to\infty} [\bar{g}(\mu)]^{-d_0/b_0} \Aren ,
\end{eqnarray*}
  which -- in contrast to $ \Aren$ renormalized at scale $\mu$ --
  is scale and scheme independent.

  Following ~\cite{CAPITANI2}, we propose to reach this goal through
  the steps:
\begin{itemize}
\item[1.]
  Define the renormalized static axial current in
  the SF scheme, i.e. through correlation
  functions in a finite space-time volume of extent $L$ with
  SF boundary conditions. The renormalization scale is $\mu=1/L$.
\item[2.]
  Compute the renormalization constant at low energy, $\mu_0 = 1/L_{0}$,
  as a function of the bare coupling.
\item[3.]  
  Compute the step scaling
  function connecting
  the currents renormalized at different $\mu$ 
  and use it to evolve the current from $\mu_0$ to the perturbative regime
  of $\mu={\rm O}(100 {\rm GeV})$.
\item[4.] 
  Use
  perturbation theory to evolve further to infinite energy to obtain $\Argi$.
\end{itemize}
Apart from 4., all steps have to be done non-perturbatively (by MC).
In the important part 2. the continuum limit can and should be taken. 
Once one arrives at 4. all dependence on the intermediate SF-scheme is
gone and the current depends only on the bare coupling as well as the
details of the discretization. 

Up to now, we have used perturbation theory to study 1. and to 
obtain the two-loop anomalous dimension in the SF scheme, which
is necessary to render higher order perturbative effects negligible in
4. We summarize these investigations in the following sections.

\section{Static approximation and the SF}
\label{SchrodStat}
   The SF is defined on a space-time cylinder
   $L^3\times L$. 
   All details as well as notation pertaining to 
   the relativistic fields are taken over from~\cite{LUSCHER1}.

   In the static limit, $m_{\rm h} \to \infty$, the relativistic
   Dirac Lagrangian is replaced
   by~\cite{EICHHILL1}:
\begin{eqnarray}
  \label{statact}
  {\cal L}_{\rm h}(x) = \heavyb(x) D_0  \heavy(x)\,,
\end{eqnarray}
   with a  static quark field, $\heavy(x)$, satisfying
\begin{eqnarray*}
   \frac{1}{2}(1+\gamma_0)\heavy=\heavy,
   \quad\frac{1}{2}(1-\gamma_0)\heavy=0,
\end{eqnarray*}
and the time component $D_0$ of the covariant derivative.
   In the time direction, the following boundary conditions are
   imposed on the heavy quark field:
\begin{eqnarray*}
   \heavy(x)|_{x_0=0}=\rho_{\rm h}(\mb{x}), \quad
   \heavyb(x)|_{x_0=L}=\bar{\rho}_{\rm h}'(\mb{x}).
\end{eqnarray*}
   In space, both the static and the relativistic quark fields are
   periodic up to a phase
   $\theta$:
\begin{eqnarray*}
   \heavy(x+L\hat{k})=e^{i\theta}\heavy(x).
\end{eqnarray*}
   For the heavy quark field, boundary fields
\begin{eqnarray*}
   \bar{\zeta}_{\rm h}(\mb{x})=\frac{\delta}{\delta\rho_{\rm h}(\mb{x})},
   \qquad
   \zeta_{\rm h}'(\mb{x})=\frac{\delta}{\delta\bar{\rho}_{\rm
       h}'(\mb{x})},
\end{eqnarray*}
are introduced as for the light quarks following~\cite{LUSCHER1}.
In particular, the derivatives are taken at $\rho_{\rm h}=\bar{\rho}_{\rm
    h}'=0$.

   The static-light axial current,
\begin{eqnarray*}
   A_0^{\text{stat}}(x)=\bar{\psi}_j(x)\gamma_0\gamma_5 \heavy(x) \,,
\end{eqnarray*}
   involves
   a relativistic anti quark field $\bar{\psi}_j$.

   The renormalization of the SF with static quarks needs to be
   discussed. Here we make the usual assumption that the SF is finite
   after adding~\cite{SYMANZIK1}
\begin{itemize}
\item local counterterms of dimension $d\leq 4$ to the bulk action
\item and surface terms composed again out of local fields, now with
  $d\leq 3$, integrated over the surfaces $x_0=0$ and $x_0=L$.
\end{itemize}
With these assumptions, we find that apart from the renormalization of
the relativistic SF~\cite{SINT1,LUSCHER1}, we need a multiplicative
renormalization of the static boundary quark fields $\bar{\zeta}_{\rm
  h}$, $\zeta_{\rm h}'$ and a mass counter\-term,
\begin{eqnarray*}
  \delta m\,\heavyb(x)\heavy(x).
\end{eqnarray*}
  Of course, the static current is renormalized multiplicatively,
\begin{eqnarray*}
  \Aren=\zastat\Astat,
\end{eqnarray*}
  not depending on the boundary conditions.

  Above, we have used \emph{continuum} notation;
  the \emph{lattice}
  action for the static quark fields, i.~e. the discretization
  of~(\ref{statact}), is chosen as in refs.~\cite{EICHHILL1,BORRELLI1}.

\section{Correlation functions and their renormalization properties}
Starting from the correlation functions,
\begin{eqnarray*}
  f_{\rm A}^{\text{stat}}(x_0) & = & -a^6\sum_{\mb{y},\mb{z}}\frac12\langle
  A_0^{\text{stat}}(x)\bar{\zeta}_{\rm h}(\mb{y})\gamma_5\zeta_i(\mb{z})
  \rangle \,,
 \\
  f_1^{\text{stat}} 
  & = & -\frac{a^{12}}{L^6}\sum_{\mb{u},\mb{v},\mb{y},\mb{z}}\frac12
  \langle\bar{\zeta}'_i(\mb{u})\gamma_5\zeta'_{\rm h}(\mb{v})\\
  & & \qquad\qquad\qquad
    \times\bar{\zeta}_{\rm h}(\mb{y})\gamma_5\zeta_i(\mb{z})\rangle,
\end{eqnarray*}
we define  
\begin{eqnarray*}
    X(u,a/L)=\left.{ f_{\rm A}^{\text{stat}}(L/2) \over
                           \sqrt{f_1^{\text{stat}}}    }
                   \right|_{\bar{g}^2(\mu=1/L)=u}.
\end{eqnarray*}
In this ratio, both the wave function renormalization constants
and the static quark mass counterterm cancel, such that $X$ is renormalized by
current renormalization only:
$
  X_{\rm R} = \zastat X.
$

As a check of the assumptions made in section~\ref{SchrodStat}, we
computed, in perturbation theory, the matching of the ratio $X_{\rm
R}$ to the corresponding quantity $Y_{\rm R}(u,a/L,z)$
defined for two relativistic
quark flavours, where the $\overline{\rm MS}$-mass of one quark
flavour is $m_{\rm R}=z/L$, and the other relativistic quark mass is
set to zero.

Defining also the finite parts of the current renormalizations in the
$\overline{\rm MS}$ scheme,
the matching condition
\begin{eqnarray}
  \label{matching}
   Y_{\rm R} = X_{\rm R}
    +{\rm O}(\frac{a}{L})+ {\rm O}(\frac{1}{z}).
\end{eqnarray}
holds. Here,
${\rm O}(\frac{1}{z})$ stands for ${\rm O}((\log{z})^n/z)$, with
$n\leq 1$ at 1-loop.

We have checked eq.(\ref{matching}) explicitly at 1-loop order,
controlling both the ${\rm O}(\frac{a}{L})$ and the
${\rm O}(\frac{1}{z})$ terms by extrapolations.
This shows that the renormalization works as
expected and at the same time 
confirms the result of~\cite{BORRELLI1} for $\zastat$
(without the need of an infrared regulator).

\section{Renormalization in the SF scheme}
In the SF scheme, the finite parts of the renormalization constants
are defined by the \emph{renormalization condition}
\begin{eqnarray*}
  X_{\rm R} = X^{(0)}\;\; \longrightarrow \quad
  \zastat=\frac{X^{(0)}}{X}.
\end{eqnarray*}
Choosing zero background field as in \cite{CAPITANI2},
the constant $\zastat$ still depends on the scale $1/L$ and on the parameter
$\theta$, which remains free. To determine the scale dependence of the
renormalization constant,
the step scaling function $\Sigma_A^{\text{stat}}$, defined by
\begin{eqnarray*}
  \zastat(2L)&=&\Sigma_{\rm A}^{\text{stat}}(u,a/L)
                   \zastat(L),\\
  u&=&\bar{g}^2(\mu=1/L),
\end{eqnarray*}
is introduced.
  The continuum limit, $\sigma(u)$, satisfies
\begin{eqnarray*}
  \Sigma(u,a/L) & = & \sigma(u) + {\rm O}(a/L), \\ 
  \sigma(u) & = & 1+ d_0\,\log{(2)}\, u + ... \,,
\end{eqnarray*}
  where $d_0$ is the  1-loop coefficient of the
  static axial current's anomalous dimension.

\begin{figure}
  \vspace{1mm}
  \epsfig{file=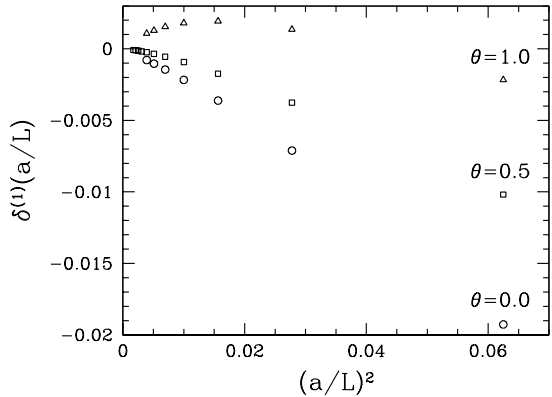,width=7.5cm,clip=}
  \vspace{-14mm}
  \caption{Discretization errors in the 
    step scaling function after O$(a)$ improvement.}
  \label{deltaplot}
\end{figure}

  As an estimate for the discretization errors,
\begin{eqnarray*}
  \delta(u,a/L) & \equiv & {\Sigma(u,a/L)- \sigma(u) \over 
                               \sigma(u)} \\
                      & = & 0 + \delta^{(1)}(a/L)u + \ldots
\end{eqnarray*}
  is computed in perturbation theory
  (figure~\ref{deltaplot}). At one-loop level, the discretization
  errors are as small as a few per cent, when $O(a)$ improvement is employed.
  We therefore expect that $\Sigma$, computed by MC-simulations, can
  be extrapolated to its continuum limit $\sigma$.

  The relation to the $\overline{\rm MS}$ scheme is given by
\begin{eqnarray*}
  (A_{\rm R,\overline{\rm MS}}^{\text{stat}})_0 & = &
  (A_{\rm R,{\rm SF}}^{\text{stat}})_0
  (1 + c_1(\theta) \bar{g}^2 + \ldots ),\\
  \mu & = & 1/L \,,
\end{eqnarray*}
  with $ c_1(\theta) $ 
  independent of the regularization.
  As an example we found
\begin{eqnarray}
  \label{sftomsbar}
    c_1(0.5) = -0.0352(2).
\end{eqnarray}
  From~(\ref{sftomsbar}) and the two-loop anomalous dimension in the
  $\overline{\rm MS}$ scheme~\cite{GIMENEZ1}, we computed
  the two-loop anomalous
  dimension in the SF scheme,
\begin{eqnarray*}
  d_1^{\text{SF}}(\theta=0.5) = 
  \frac{1}{(4\pi)^2}\{0.066(4)-0.0455(3)\,N_{\rm f}\},
\end{eqnarray*}
  with $N_{\rm f}$ relativistic quark flavours.

\section{Concluding remarks}
  We have defined a renormalization scheme, which, through steps
1.-4., should allow 
 to solve the renormalization problem
for the  static axial current. Cutoff effects in the associated 
step scaling function and the relation to other schemes have been
computed to one-loop but the   
signals in Monte Carlo computations remain to be investigated.

We finally remark that it may be advantageous to
also compute the decay constant $f_{\rm B}^{\text{stat}}$
using the ratio $X$, but now 
  for large values of the
  time-extent of the SF~\cite{GUAGNELLI1}.

\end{document}